\documentstyle[11pt,epsfig]{article}
\begin{document}
\title{\Large\bf  On the nature of the $\pi_2(1880)$}

\author{\small De-Min Li\footnote{E-mail: lidm@zzu.edu.cn}~~and Shan Zhou\\
\small   Department of Physics, Zhengzhou University, Zhengzhou,
Henan 450052, People's Republic of China}
\date{\today}
\maketitle
\vspace{0.5cm}

\begin{abstract}

The strong decays of the $\pi_2(1880)$ as the $2\,^1D_2$
quark-antiquark state are investigated in the $^3P_0$ model and
the flux-tube model, respectively. The results are similar in the
two models. It is found that the decay patterns of the
conventional $2\,^1D_2$ meson and the $2^{-+}$ light hybrid are
very different, and the experimental evidence for the
$\pi_2(1880)$ is consistent with it being the conventional
$2\,^1D_2$ meson rather than the $2^{-+}$ light hybrid. The
possibility of the $\pi_2(1880)$ being a mixture of the
conventional $q\bar{q}$ and the hybrid is discussed.

\end{abstract}

\vspace{0.5cm}


\newpage

\baselineskip 24pt

\section*{I. Introduction}
\indent \vspace*{-1cm}

Experimentally, the ACCMOR Collaboration in 1981 observed a $I=1$,
$J^{PC}=2^{-+}$ structure at 1850 MeV in the $f_2(1270)\pi$
$D$-wave
 with a width of about 240 MeV\cite{accmor}. Subsequently, the VES
Collaboration reported a $J^{PC}=2^{-+}$ threshold enhancement in
the $a_2(1320)\eta$ channel with a mass of about 1840 MeV and
width of about 210 MeV in their $\eta\eta\pi^-$ data\cite{ves1996}
and also in $\eta\pi^+\pi^-\pi^0$ data where
$a_2(1320)\rightarrow\pi^+\pi^-\pi^0$; they also observed a strong
peak at the same mass in the $f_2(1270)\pi$ $D$-wave in their
$4\pi$ data\cite{ves1999}. In 2001, Anisovich et al. reported a
$I=1$, $J^{PC}=2^{-+}$ resonance with a mass of about 1880 MeV and
a width of about 255 MeV in the $a_2(1320)\eta$ and $f_2(1270)\pi$
$D$-wave\cite{anisovich2001,anisovich20012}. More recently, a
similar $2^{-+}$ resonance was observed by the E852 Collaboration
in the $f_1(1285)\pi$\cite{e8521}, $\rho\omega$\cite{e8522}, and
$a_2(1320)\eta$\cite{e8523} channels, respectively. It has been
established that these observations in different channels refer to
a single state $\pi_2(1880)$\cite{bugg, pdg08}. In the Meson
Summary Table of the PDG2008, the mass and width of the
$\pi_2(1880)$ are quoted to be $1895\pm 16$ MeV and $235\pm 34$
MeV, respectively\cite{pdg08}.

As for the $\pi_2(1880)$ nature, after a $2^{-+}$ hybrid
conjecture [$\pi_2(H)$] was first proposed by Anisovich et
al.\cite{anisovich2001}, several groups also claimed the
$\pi_2(1880)$ being a viable non-exotic hybrid
candidate\cite{anisovich20012,e8521,e8522,e8523,bugg,klempt}. With
the $\pi_2(1670)$ as the well-established $1\,^1D_2$ $q\bar{q}$
state\cite{pdg08}, the $\pi_2(1880)$ looks like the $\pi_2(H)$
rather than the $2\,^1D_2$ isovector $q\bar{q}$ state
[$\pi_2(2D)]$ based on its mass, because the observed mass of the
$\pi_2(1880)$ just overlaps the flux-tube model prediction of
about $1.8\sim 1.9$ GeV for the light $2^{-+}$
hybrid\cite{ftmass}, but is about 200 MeV lower than the
Godfrey-Isgur (GI) quark model prediction of about 2.1 GeV for the
$2\,^1D_2$ nonstrange $q\bar{q}$ state\cite{GI}. However,
comparing the experimental evidence for the $\pi_2(1880)$ with the
strong decay properties of the $\pi_2(H)$ expected by the model of
hybrid meson decay developed by Page, Swanson, and
Szczepaniak(PSS) based on the heavy quark expansion of QCD and the
strong coupling flux tube picture of nonperturbative
glue\cite{pss}(see the Table II of Ref.\cite{pss}), one can find
the following features of the $\pi_2(1880)$ casting doubt over the
hybrid interpretation for the $\pi_2(1880)$:

i) The observation in the $\rho\omega$ channel of the
$\pi_2(1880)$ is inconsistent with the hybrid interpretation where
the coupling of the $\pi_2(H)$ to $\rho\omega$ is expected to
vanish.

ii) The observation in the $f_2(1270)\pi$ $D$-wave of the
$\pi(1880)$ is also inconsistent with the hybrid interpretation
where the $f_2(1270)\pi$ $D$-wave is strongly suppressed and the
$S$-wave is dominant.

iii) The measured width of the $\pi_2(1880)$, $235\pm 34$ MeV, is
again inconsistent with the hybrid interpretation where the
$\pi_2(H)$ width is less than 100 MeV.

Therefore, the claims for the $\pi_2(1880)$ can be accepted as a
resonance state of exotic nature may be premature. In fact, it is
important to exhaust possible conventional $q\bar{q}$ description
of the $\pi_2(1880)$ before resorting to more exotic
interpretation such as a hybrid. In this work, we shall discuss
the possibility of the $\pi_2(1880)$ being the $\pi_2(2D)$. As
mentioned above, a problem with identifying the $\pi_2(1880)$ with
the $\pi_2(2D)$ is that its mass is about 200 MeV lower than the
expectation from the GI quark model. Notice that the $a_1(1700)$
and $a_2(1700)$, both about 100-200 MeV lower in mass than the GI
quark model anticipated\cite{GI}, turn out the excellent
candidates for radial excitations\cite{pss,3p02}, which indicates
that GI quark model maybe overestimate the masses of the
higher-$L$ radially excited mesons by about 100-200 MeV, and
therefore the $\pi_2(2D)$ with a mass about 1.9 GeV is not
implausible.  Also, the mass of the $\pi_2(2D)$ in the spectrum
integral equation\cite{spectrum} is expected to be about 1.937
GeV, very close to the $\pi_2(1880)$ mass. Therefore, the
assignment of the $\pi_2(1880)$ as the $\pi_2(2D)$ seems also
possible based on its mass. However, only the $\pi_2(1880)$ mass
information is insufficient to identify its nature, further
studies of its decay dynamics are needed. The main purpose of this
work is to discuss whether the $\pi_2(2D)$ interpretation for the
$\pi_2(1880)$ is reasonable or not by investigating its strong
decay properties in two models, the $^3P_0$ model and the
flux-tube model.

The organization of this paper is as follows. In Sec. II, the
decay properties of the $\pi_2(1880)$ as the $\pi_2(2D)$ within
the $^3P_0$ model and the flux-tube model are presented. The
discussions are presented in Sec. III, and the summary and
conclusion are given in Secs. IV.

\section*{II. Decay properties of the $\pi_2(1880)$ as the $\pi_2(2D)$}
\indent\vspace{-1cm}

The $^3P_0$ model and the flux-tube model which are the standard
models for strong decays at least for mesons in the initial state,
have been widely used to evaluate the strong decays of
hadrons\cite{3p02,3p0rev1,3p0rev2,3p0rev3,3p0rev4,flux,3p0y,3p0x,3p01,3p03,quarkmass,3p04},
since they give a good description of many of the observed decay
amplitudes and partial widths of the hadrons. In this work, we
shall employ the $^3P_0$ model and the flux-tube model with simple
harmonic oscillator (SHO) wave functions\footnote{This is typical
of decay calculations and it has been demonstrated that using the
more realistic wave functions, such as those obtained from
Coulomb, plus the linear potential model, does not change the
results significantly\cite{flux,3p0y,3p0x}.} to evaluate the
two-body open-flavor strong decay widths of the initial state.
Since there exists exhaustive literature on these two models, we
just list the numerical values of the partial decay widths of the
$\pi_2(1880)$ as the $\pi_2(2D)$ in Table 1.  In our calculation,
the SHO wave function scale parameter $\beta$, the pair production
strength parameter $\gamma$ in the $^3P_0$ model, the
pair-creation constant $\gamma_0$ and the string tension $b$ in
the flux-tube model, and the constituent quark mass $m_q$
are\footnote{Our value of $\gamma$ is higher than that used by
Ref.\cite{quarkmass} (0.505) by a factor of $\sqrt{96\pi}$ due to
different field conventions, constant factor in $T$, etc. The
calculated results of the widths are, of course, unaffected.}
$\beta_A=\beta_B=\beta_C=\beta=0.4$ GeV, $\gamma=8.77$,
$\gamma_0=14.3$, $b=0.18$ GeV$^2$, $m_u=m_d=0.33$ GeV, and
$m_s=0.55$ GeV\cite{3p0rev4,quarkmass}, respectively. The meson
masses used to determine the phase space and final state momenta
are\footnote{ We assume that the $f_0(1370)$ is the ground scalar
meson as Refs.\cite{3p02,3p01,3p03}.} $M_{\pi}=138$ MeV ,
$M_K=496$ MeV, $M_{\eta}=548$ MeV, $M_{\rho}=776$ MeV,
$M_{K^\ast}=894$ MeV, $M_{\omega}=783$ MeV, $M_{a_1(1230)}=1230$
MeV, $M_{f_1(1285)}=1282$ MeV, $M_{a_2(1320)}=1318$ MeV,
$M_{f_2(1270)}=1275$ MeV, $M_{f_0(1370)}=1370$ MeV,
$M_{\rho(1450)}=1465$ MeV, and $M_{K_1(1270)}=1272$ MeV.

{\small
\begin{table}[hbt]
\begin{center}
\vspace*{-0.5cm}
 \caption{\small Decays of the $\pi_2(1880)$ as the
$\pi_2(2D)$ in the $^3P_0$ model and the flux-tube model. The
initial state mass is set to $1895$ MeV.}
 \vspace*{0.5cm}

\begin{tabular}{c|c|c}\hline\hline
Mode  & $\Gamma_{LS}$ in $^3P_0$ model (MeV) & $\Gamma_{LS}$  in
flux-tube model (MeV)
\\\hline

$\rho\pi$   & $\Gamma_{P1}=5.58$  &  $\Gamma_{P1}=6.07$ \\
            & $\Gamma_{F1}=66.04$  &$\Gamma_{F1}=71.92$  \\\hline
$K^\ast K$  &  $\Gamma_{P1}=14.46$  & $\Gamma_{P1}=15.75$\\
             & $\Gamma_{F1}=5.76$  &$\Gamma_{F1}=6.27$   \\\hline
$\rho\omega$&  $\Gamma_{P1}=29.24$  &$\Gamma_{P1}=31.84$\\
 & $\Gamma_{F1}=10.19$  &  $\Gamma_{F1}=11.10$\\\hline
$K^\ast K^\ast$ & $\Gamma_{P1}=5.25$     & $\Gamma_{P1}=5.71$   \\
 & $\Gamma_{F1}=0.06$  & $\Gamma_{F1}=0.07$ \\\hline
$\rho(1450)\pi$& $\Gamma_{P1}=19.15$ &$\Gamma_{P1}=11.82$\\
 & $\Gamma_{F1}=1.48$  &  $\Gamma_{F1}=1.09$\\\hline
$ f_0(1370)\pi$&$\Gamma_{D0}=4.20$ &$\Gamma_{D0}=4.57$\\\hline

$f_1(1285)\pi$ &$\Gamma_{D1}=5.29$ &$\Gamma_{D1}=5.76$\\\hline

$a_1(1260)\eta$& $\Gamma_{D1}=0.36$&$\Gamma_{D1}=0.40$\\\hline

$ K_1(1270)K$ &$\Gamma_{D1}=0.25$ &$\Gamma_{D1}=0.27$\\\hline

$a_2(1320)\eta$ &$\Gamma_{S2}=20.86$ &$\Gamma_{S2}=22.71$\\
 & $\Gamma_{D2}=0.05$  & $\Gamma_{D2}=0.06$  \\
 &$\Gamma_{G2}=0.00$&$\Gamma_{G2}=0.00$\\\hline
$ f_2(1270)\pi$ &$\Gamma_{S2}=11.82$ &$\Gamma_{S2}=12.88$\\
                 &$\Gamma_{D2}=22.58$&$\Gamma_{D2}=24.59$\\
                 &$\Gamma_{G2}=0.57$&$\Gamma_{G2}=0.62$\\\hline
$\Gamma$     & 223.19       &233.50\\\hline\hline
\end{tabular}
\end{center}
\end{table}
}

 It is clear from Table 1 that the numerical results in the $^3P_0$
model are similar to those in the flux-tube model. Very
characteristic differences between the $\pi_2(2D)$ and $\pi_2(H)$
assignments for the $\pi_2(1880)$ are evident when we compare our
results with the expectations from the PSS model for the
$\pi_2(H)$\cite{pss}. The total width of the $\pi_2(2D)$ is
expected to be about 223 MeV in the $^3P_0$ model or about 233 MeV
in the flux-tube model, both in good agreement with the
$\pi_2(1880)$ width; however, the total width of the $\pi_2(H)$ is
expected to be less than 100 MeV, at least 100 MeV lower than the
experiment. The partial width of the $\pi_2(2D)\rightarrow
\rho\omega$ is significantly large, consistent with the
observation in the $\rho\omega$ channel of the $\pi_2(1880)$;
whereas the $\pi_2(H)\rightarrow\rho\omega$ is expected to vanish.
The $\pi_2(2D)\rightarrow f_2(1270)\pi$ is dominant in the
$D$-wave and the $D$-wave width is significantly large, and
therefore the $\pi_2(2D)$ should be readily observable in the
$f_2(1270)\pi$ $D$-wave, consistent with the observation of the
$\pi_2(1880)$ in the $f_2(1270)\pi$ $D$-wave; while the
$\pi_2(H)\rightarrow f_2(1270)\pi$ is strongly suppressed in the
$D$-wave and dominant in the $S$-wave. Also, for the $\pi_2(2D)$,
the partial width of the $K^\ast K^\ast$ mode, the $F$-wave widths
of the $\rho\pi, K^\ast K, K^\ast K^\ast$ and $\rho(1450)\pi$
modes, and the $G$-wave width of the $f_2(1270)\pi$
 mode are not zero, especially the $\rho\pi$ $F$-wave width is
significantly large; whereas for the $\pi_2(H)$, all these widths
vanish exactly. The further experimental study on these decay
modes are also important to examine whether the $\pi_2(1880)$ is
the $\pi_2(2D)$ or the $\pi_2(H)$.

From these remarkable discriminants between the $\pi_2(2D)$ and
the $\pi_2(H)$, it is clear that the available experimental
evidence for the $\pi_2(1880)$ is consistent with it being the
$\pi_2(2D)$ rather than the $\pi_2(H)$\footnote{The one exception
to this is that our predicted $\Gamma(a_2(1320)\eta)/\Gamma(
f_1(1285)\pi)$ for the $\pi_2(2D)$ is about 4, inconsistent with
the measured value of $22.7\pm 7.3$\cite{e8521} which agrees with
the PSS model prediction of about 23 for the
$\pi_2(2H)$\cite{pss}. Notice that the systematic error is not
estimated in this measured datum, and this datum is not used for
averages, fits, limits, etc. by PDG2008\cite{pdg08}. The further
confirmation of this ratio is needed.}, assuming the $^3P_0$ model
and the flux-tube model are accurate.

 In order to test the robustness of our
results, the dependence of the predicted results on the initial
state mass $M_A$ and the SHO function scale parameter $\beta$ is
studied. We show the variation of the total width of the
$\pi_2(2D)$ with $M_A$ and $\beta$ in Fig. 1. In both the $^3P_0$
model and the flux-tube model, the total width of the $\pi_2(2D)$
becomes large with the increase of the $M_A$, and it always lies
in the width range of the $\pi_2(1880)$. When the $\beta$ varies
from 300 to 500 MeV, in the flux-tube model the $\pi_2(2D)$ width
varies from about 230 to 260 MeV, depending weakly on the $\beta$,
while in the $^3P_0$ model it varies dramatically with the
$\beta$. In order to reproduce the $\pi_2(1880)$ width in the
$^3P_0$ model, it requires $\beta\simeq 370\sim 420$ MeV,
overlapping the typical value of about $350\sim 450$ MeV used in
the computation of the light meson decays for the SHO wave
functions with a common $\beta$\cite{3p0x,3p01}.

\begin{figure}[hbt]
\begin{center}
\epsfig{file=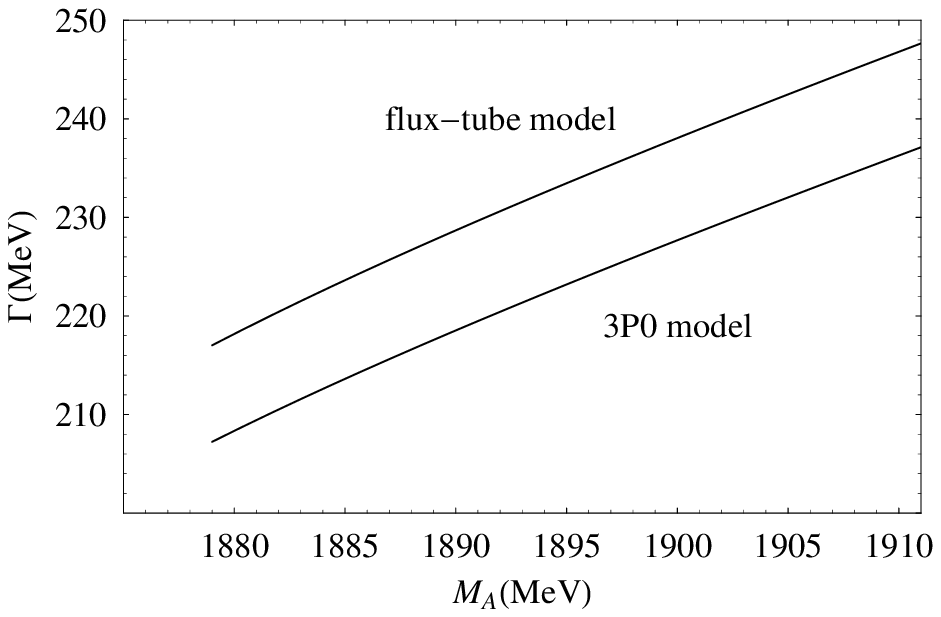,width=6.0cm, clip=}
\epsfig{file=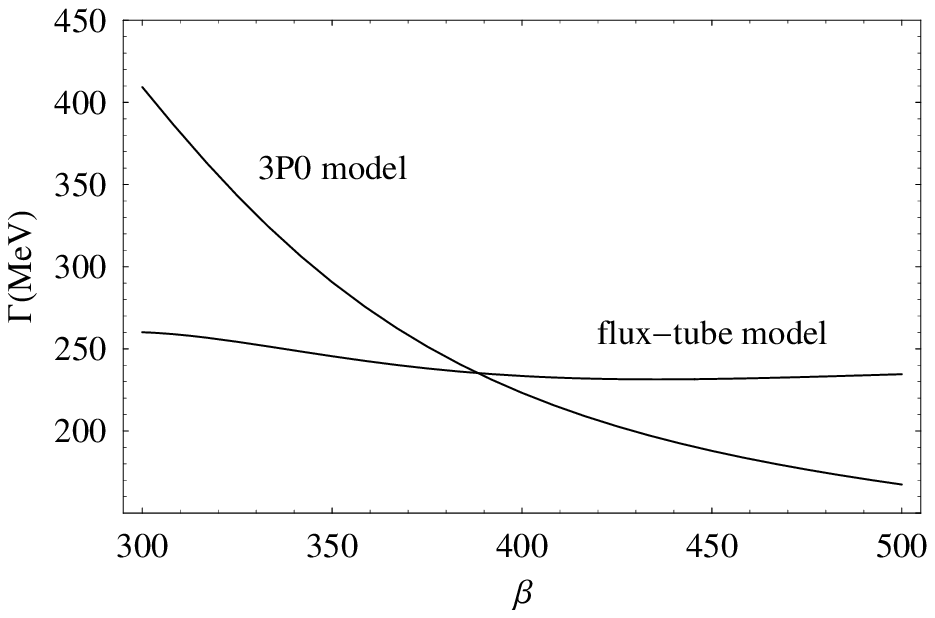,width=6.0cm, clip=}
\vspace*{0.5cm}\vspace*{-0.8cm}
 \caption{\small The total width of the $\pi_2(2D)$ dependence on
the $M_A$ and $\beta$ in the $^3P_0$ model and the flux-tube
model.}
\end{center}
\end{figure}

\begin{figure}[hbt]
\begin{center}
\epsfig{file=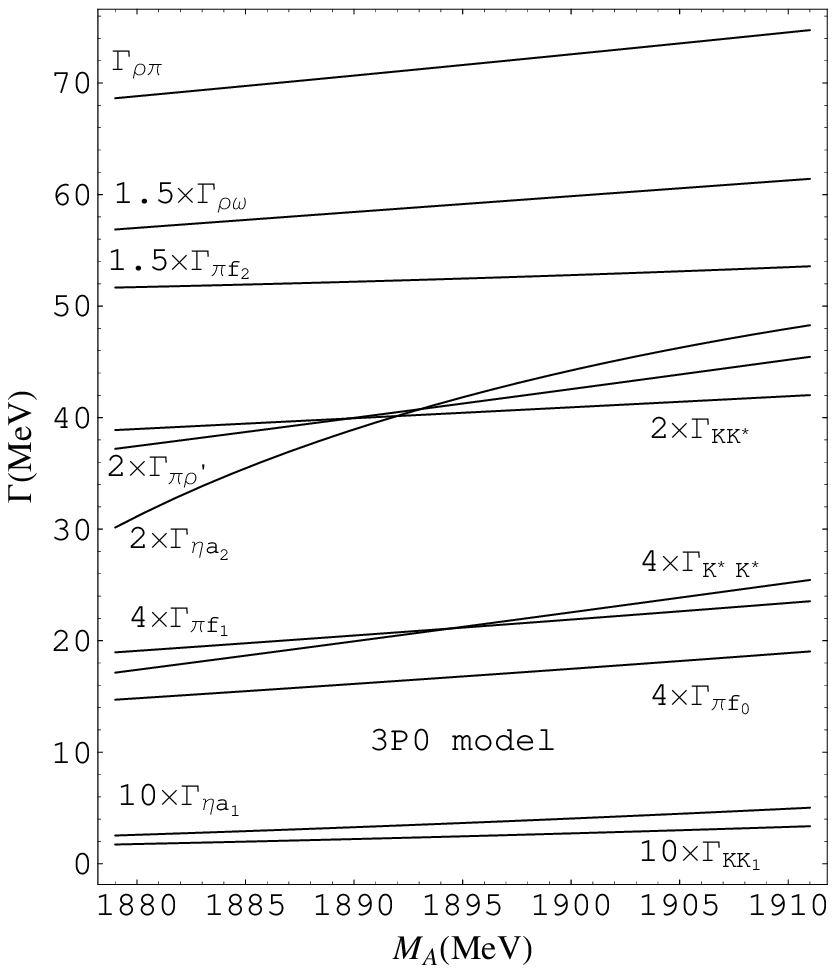,width=6.0cm, clip=}
\epsfig{file=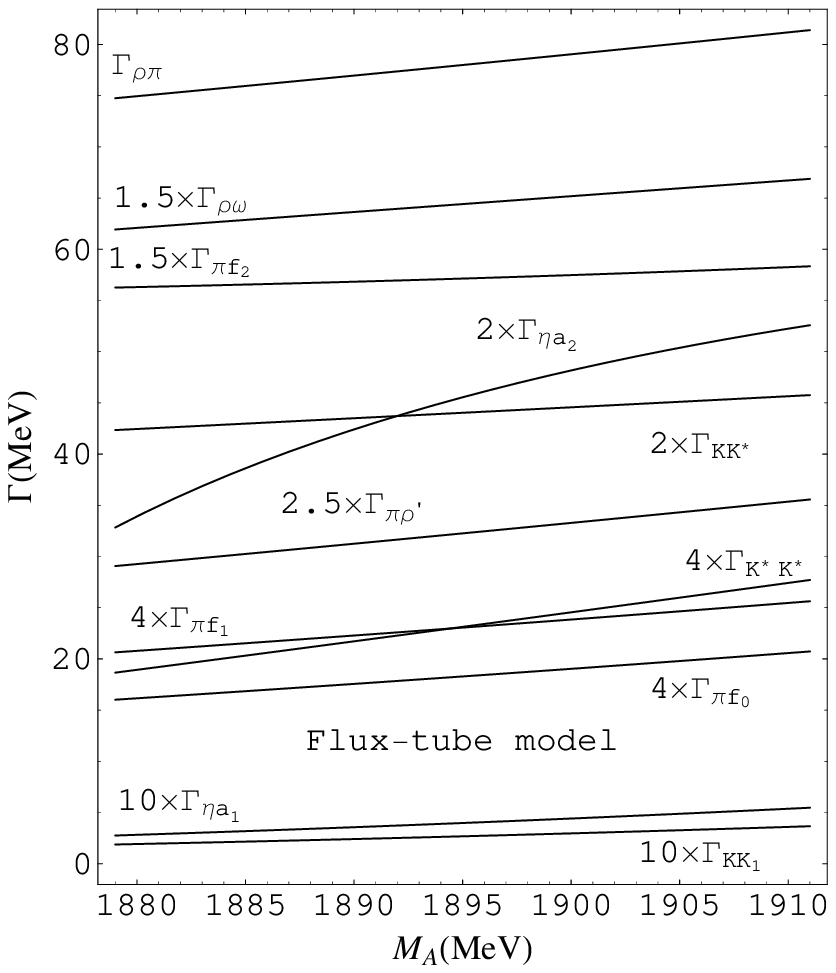,width=6.0cm, clip=} \caption{\small The
partial widths of the $\pi_2(2D)$ dependence on
 the $M_A$ in the $^3P_0$ model and the flux-tube model. $\rho^\prime$ denotes the $\rho(1450)$.}
\end{center}
\end{figure}

The partial widths of the $\pi_2(2D)$ versus the $M_A$ are shown
in Fig. 2, where the variations of the partial widths of the
$\pi_2(2D)$ with the $M_A$ in the $^3P_0$ model are similar to
those in the flux-tube model. The partial widths increase when the
$M_A$ increases and the dominant decay modes are $\rho\pi$,
$\rho\omega$, $f_2(1270)\pi$, $KK^\ast$, $\rho(1450)\pi$ and
$a_2(1320)\eta$.

\begin{figure}[hbt]
\begin{center}
\epsfig{file=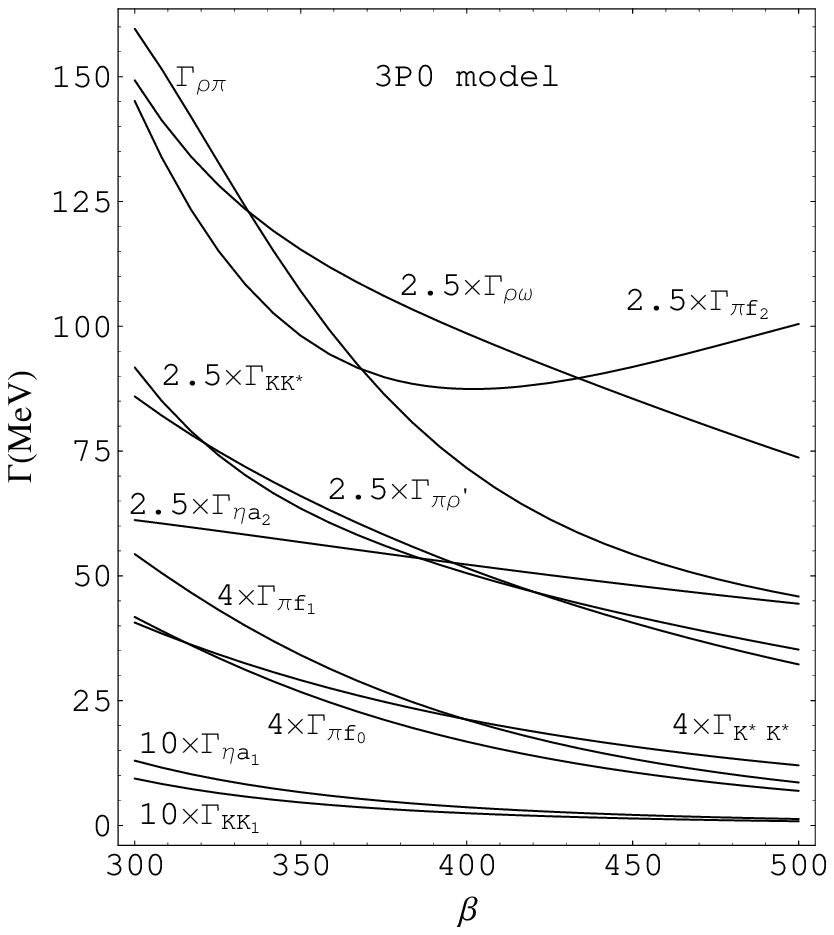,width=6.0cm, clip=}
\epsfig{file=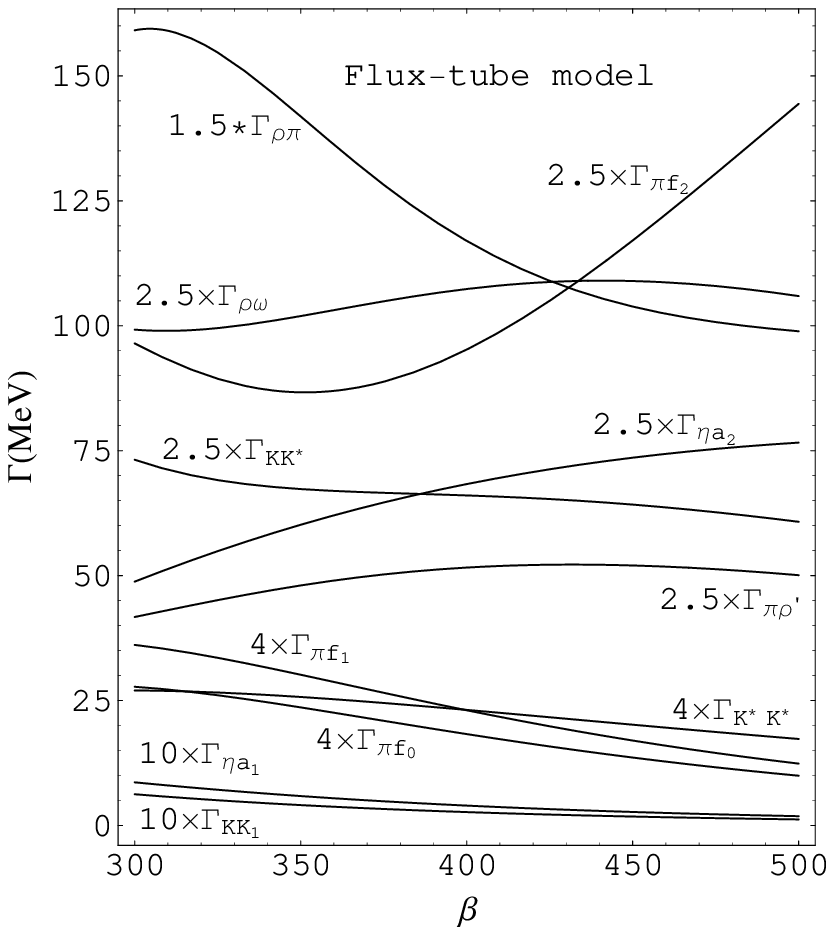,width=6.0cm, clip=} \caption{\small The
partial widths of the $\pi_2(2D)$ dependence on
 the $\beta$ in the $^3P_0$ model and the flux-tube model. $\rho^\prime$ denotes the $\rho(1450)$.}
\end{center}
\end{figure}

The partial widths of the $\pi_2(2D)$ versus the $\beta$ are shown
in Fig. 3. We see from Fig. 3 that in two models, $\rho\pi$,
$\rho\omega$, $f_2(1270)\pi$, $KK^\ast$, $\rho(1450)\pi$ and
$a_2(1320)\eta$ are still the dominant decay modes when the
$\beta$ varies. For small $\beta$ ($\beta\simeq 300\sim 350$ MeV),
$KK^\ast$ dominates $a_2(1320)\eta$, however for large $\beta$
($\beta\simeq 450\sim 500$ MeV), $a_2(1320)\eta$ dominates
$KK^\ast$. In the vicinity of $\beta=400$ MeV,
$\Gamma(KK^\ast)\simeq \Gamma(a_2(1320)\eta)$. The similar
behavior also exists for the modes $f_1(1285)\pi$ and $K^\ast
K^\ast$. The measurement of the
$\Gamma(KK^\ast)/\Gamma(a_2(1320)\eta)$ and
$\Gamma(f_1(1280)\pi)/\Gamma(K^\ast K^\ast)$ for the $\pi_2(1880)$
would be useful for the reasonable choice for the $\beta$.

We note that for the $\pi_2(2D)$, the $f_2(1270)\pi$ $D$-wave
dominates the $S$-wave and the $\rho\pi$ $F$-wave dominates the
$P$-wave, which is unusual because in most cases the lower partial
waves are dominant. As mentioned above, these results are
remarkably different with the expectations from the PSS
model\cite{pss} for the $\pi_2(H)$. We find that in both the
$^3P_0$ model and the flux-tube model, the $F$-wave always
dominates the $P$-wave for the $\pi_2(2D)\rightarrow\rho\pi$ and
the $D$-wave always dominates the $S$-wave for the
$\pi_2(2D)\rightarrow f_2(1270)\pi$ when the $M_A$ varies in the
mass range of the $\pi_2(1880)$ and $\beta$ varies in the range
370-420 MeV, the required range for reproducing the $\pi_2(1880)$
width in the $^3P_0$ model mentioned above. Determining these
partial widths ratios experimentally is very important to
distinguish the $\pi_2(2D)$ interpretation from the $\pi_2(H)$
assignment for the $\pi_2(1880)$.

\section*{III. Discussions}
\indent\vspace{-1cm}

Generally speaking, the pure $\pi_2(2D)$ can mix with the pure
$\pi_2(H)$ to produce the physical state $\pi_2(1880)$. The
available experimental evidence for the $\pi_2(1880)$ is in favor
of the $\pi_2(2D)$ interpretation for the $\pi_2(1880)$ based on
the remarkably different decay patterns of the $\pi_2(2D)$ and
$\pi_2(H)$, but it is insufficient to quantitatively determine the
$q\bar{q}$-hybrid content of the $\pi_2(1880)$\footnote{Within the
$\pi_2(1880)$ being the mixture of the $\pi_2(2D)$ and $\pi_2(H)$,
the measured partial widths of the $\pi_2(1880)$ are needed to
determine the hybrid-quarkonium content of the $\pi_2(1880)$
quantitatively.}, which is essential to confirm or refute that the
possibility of the hybrid admixture in the $\pi_2(1880)$.
Therefore, the possibility of the $\pi_2(1880)$ being in fact a
mixture of the $\pi_2(2D)$ and $\pi_2(H)$ might exist at present
time.

We can qualitatively estimate the hybrid component of the
$\pi_2(1880)$ would be small based on its available experimental
information. The $\pi_2(H)\rightarrow \rho\omega$ is expected to
vanish from the PSS model\cite{pss}, therefore, the observation of
the $\pi_2(1880)$ in the $\rho\omega$ channel\cite{e8522} makes
that the substantial hybrid admixture in the $\pi_2(1880)$ seems
impossible. However, it should be noted that the $\pi_2(1880)$
signal in the $\rho\omega$ channel was observed only by the E852
Collaboration\cite{e8522}, and even it is not clear whether the
unitarity conserving fit to the $\rho\omega$ mass distributions
would need to have the $\pi_2(1880)$ decaying to $\rho\omega$ or
not\footnote{We thank the anonymous referee for pointing out this
matter.}. Therefore, further evidence is needed to confirm whether
the hybrid component of the $\pi_2(1880)$ is small or not.
Fortunately, as mentioned in Sec. I, the $\pi_2(1880)$ has been
observed by three different groups in the $f_2(1270)\pi$ $D$-wave,
which implies that the $\pi_2(H)$ component of the $\pi_2(1880)$
would be small because the $\pi_2(H)\rightarrow f_2(1270)\pi$ is
strongly suppressed in the $D$-wave. Similarly, the further
experimental information of the $\pi_2(1880)$ in the $K^\ast
K^\ast$ and $[\rho\pi]_{L=3}$ channels would be useful to shed
light on this issue.

Finally, our predicted $\Gamma(a_2(1320)\eta)/\Gamma(
f_1(1285)\pi)$ for the $\pi_2(2D)$ inconsistent with the
measurement of the E852 Collaboration\cite{e8521} may be a hint
for the $\pi_2(1880)$ being in fact a mixture of the $\pi_2(2D)$
and $\pi_2(H)$, and the small hybrid admixture in the
$\pi_2(1880)$ might make this measured ratio shift from the
predicted value for the pure $\pi_2(2D)$.

\section* {IV. Summary and conclusion}
\indent \vspace*{-1cm}

The strong decays of the $\pi_2(1880)$ as the $\pi_2(2D)$ are
investigated in both the $^3P_0$ model and the flux-tube model.
The overall behaviors of the decay modes in the $^3P_0$ model are
similar to those in the flux-tube model. The decay properties of
the $\pi_2(2D)$ and the $\pi_2(H)$ are remarkably different. The
decay modes $\rho\pi$, $\rho\omega$, $f_2(1270)\pi$, $K^\ast K$,
$K^\ast K^\ast$ and $\rho(1450)\pi$ are crucial for distinguishing
the conventional quarkonium interpretation from the hybrid
assignment of the $\pi_2(1880)$. The available experimental
evidence for the $\pi_2(1880)$ is consistent with it being the
conventional $2\,^1D_2$ $q\bar{q}$ meson rather than the light
$2^{-+}$ hybrid. The possibility of the small hybrid admixture in
the $\pi_2(1880)$ might exist. Further experimental study on the
partial widths of the $\pi_2(1880)$ is desirable. We tend to
conclude that the $\pi_2(1880)$ is the convincing $2\,^1D_2$
$q\bar{q}$ state or the $2\,^1D_2$ $q\bar{q}$ with small hybrid
admixture.

 \section*{Acknowledgments}
 This work
is supported in part by HANCET under Contract No. 2006HANCET-02,
and by the Program for Youthful Teachers in University of Henan
Province.

\end{document}